\documentclass[aps,prl,twocolumn,showpacs,superscriptaddress,floatfix]{revtex4}
\usepackage{graphicx}
\usepackage{dcolumn}
\usepackage{amsfonts}
\usepackage{amsmath}
\begin{document}

\title{Microscopic Study of the  triple-$\alpha$ Reaction}


\author{A. S. Umar}
\affiliation{Department of Physics and Astronomy, Vanderbilt University, Nashville, Tennessee 37235, USA}
\author{J. A. Maruhn}
\affiliation{Institut f\"ur Theoretische Physik, Goethe-Universit\"at, D-60438 Frankfurt am Main, Germany}
\author{N. Itagaki}
\affiliation{Department of Physics, University of Tokyo, Hongo, Tokyo 113-0033, Japan}
\author{V. E. Oberacker}
\affiliation{Department of Physics and Astronomy, Vanderbilt University, Nashville, Tennessee 37235, USA}

\date{\today}


\begin{abstract}
We present a microscopic dynamical study of the reactions involving three $^{4}$He clusters.
We show that the much discussed triple-$\alpha$ linear chain configuration of $^{12}$C  is formed with a certain
lifetime and subsequently decays into a triangular configuration of $^{12}$C and then to a configuration
near the ground state.  Time-dependent
Hartree-Fock (TDHF) theory coupled with a density constraint is used to study the properties
of these configurations.
We find a sequence of dynamical transitions analogous to the suggested astrophysical  mechanism for the
formation of $^{12}$C nuclei.
\end{abstract}
\pacs{21.60.-n,21.60.Jz,21.30.Fe,21.60.Cs,27.20.+n,27.30.+t1}
\maketitle

Reactions involving three $^{4}$He nuclei carry both astrophysical and
nuclear structure significance. Astrophysically, the triple-$\alpha$ reaction
is believed to be one of the dominant helium burning reactions leading
to the synthesis of most of the $^{12}$C found in the universe~\cite{WF86,SA05}.
In the first stage of this process two helium
nuclei combine to form an unstable $^{8}$Be nucleus, which has an
extremely short lifetime ($\sim 10^{-16}$s) and decays into two $\alpha$'s. However,
for densities attained during the helium burning process there will
be a small abundance of $^{8}$Be present to allow the combination
with another $^{4}$He nucleus to form the well known~\cite{Ho54} $0_{2}^{+}$ resonant state of
$^{12}$C at $7.654$~MeV, which is at an energy close to the combined energies of
the reacting $^{8}$Be and $^{4}$He nuclei, thus facilitating this reaction~\cite{SA05}.
This is the reason why despite its non-existence under natural conditions $^{8}$Be plays a crucial
role in nuclear astrophysics.
The formed resonant state of $^{12}$C is believed to decay to the $2^{+}$ state
of $^{12}$C at $4.43$~MeV and subsequently to the ground state via
$\gamma$ radiation. Recently, there has also been interest in non-resonant reaction
of three alpha's as an alternate mechanism at lower temperatures~\cite{OK09}.

The exotic structure of the $0_{2}^{+}$ state of $^{12}$C and its possible association with the
mysterious $0^{+}$ state of $^{16}$O at $6.06$~MeV has also been of great interest in nuclear
structure models of light nuclei.
In 1950's  it has been conjectured that the $0_{2}^{+}$ state might be linear chain configuration
of 3 alpha's~\cite{Mo56}, however later
it has been interpreted to be  gas-like, consisting of a mixture of various configurations
of three alpha's including the linear chain configuration~\cite{YF80}.
Therefore, search for the linear chain configurations of alpha's has been performed in different places on the nuclear chart.
Suggestions that the lifetime of the linear chain configuration may be extended by the addition of extra
neutrons have led to the discussion of cluster structures in neutron-rich Be~\cite{WO96,IO00,II08}
and C isotopes~\cite{WO97,NI04,WO04}.
The search has also been extended to other light $N=Z$ nuclei~\cite{PC61,HH72,AH92,JZ94}.
However,  until now there has been no theoretical discussion on the dynamics of the linear chain state;
e.g. how the chain state of three alpha's plays a role in the early stage of the synthesis of $^{12}$C
or what are the characteristic modes of vibration of the three alpha chain.
Another problem is that,
until now most of the theoretical analyses for the cluster structures have been performed using
effective interactions, which are determined such as to reproduce the binding energies and scattering phase shifts of the clusters.
It is highly desirable to study the presence of cluster configurations in a more general manner without the a priori
initialization in terms of clusters.
Also, dynamical oscillation of the linear chain configuration is a completely new approach to the problem.
In our previous study, cluster configurations of neutron-rich C isotopes have been obtained as local minima
of static Hartree-Fock (HF) solutions with axial symmetry,
and TDHF calculations were performed to examine their stability
against  quadrupole and octupole distortions~\cite{MK06,ML10}.
The linear chain configurations were found to be relatively stable with respect to the breathing (quadrupole) distortions
while they were unstable with respect to bending (octupole) perturbation.

It is generally acknowledged that the TDHF theory provides a
useful foundation for a fully microscopic many-body theory of low-energy heavy-ion reactions
\cite{Ne82}. Earlier TDHF calculations of collisions between light nuclei involving cluster-like
configurations have been made for the study of nuclear molecular resonances~\cite{US85}. However,
due to the lack of computational resources these calculations suffered from numerical imprecision
as well as unphysical symmetry assumptions, such as collisions being restricted to axial symmetry.
Current TDHF calculations are performed with high numerical precision and with no symmetry
assumptions as well as using modern Skyrme forces.
Recently, we have shown that when TDHF is combined with the density-constraint method~\cite{CR85} dynamical potentials and ion-ion interaction barriers can be accurately reproduced~\cite{UO06b,UO08a,UO09b}.
Thus an alternative approach would be to investigate the formation and stability of the linear chain configuration using the
fully microscopic and dynamical TDHF theory.

In this work we study TDHF collisions which reproduce the linear chain configuration and subsequently
decay to lower-energy configurations of the system. To our knowledge such mode changes have never been
observed in TDHF calculations previously and appear to simulate the suggested astrophysical  mechanism for the
formation of $^{12}$C nuclei.
\begin{figure}[!htb]
\includegraphics*[scale=0.50]{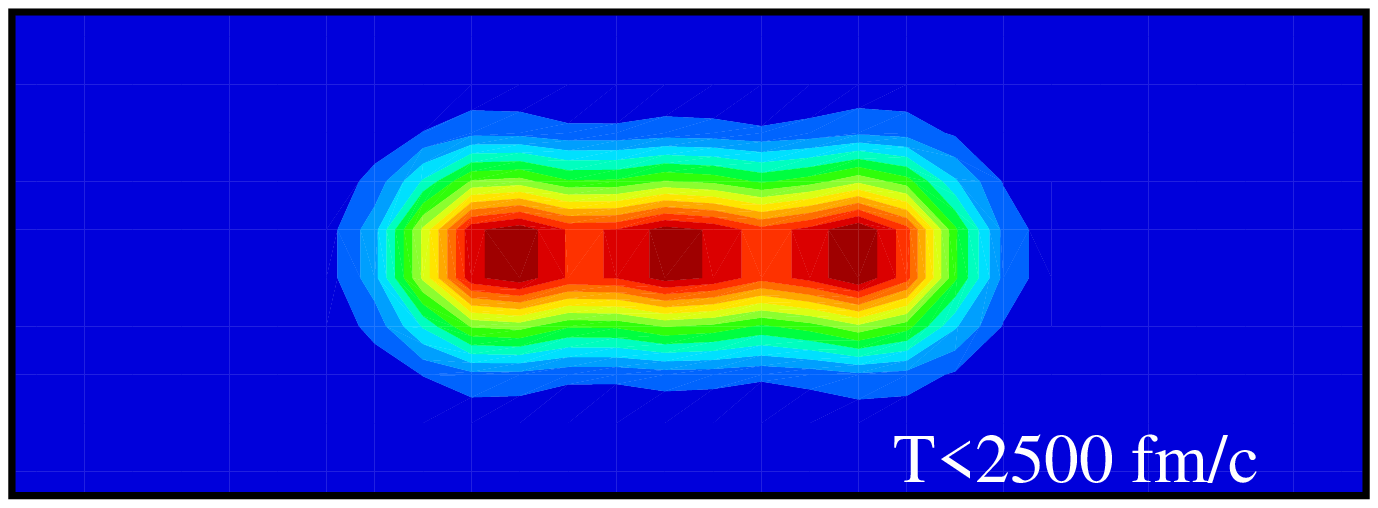}\\ \vspace{-0.03in}
\includegraphics*[scale=0.50]{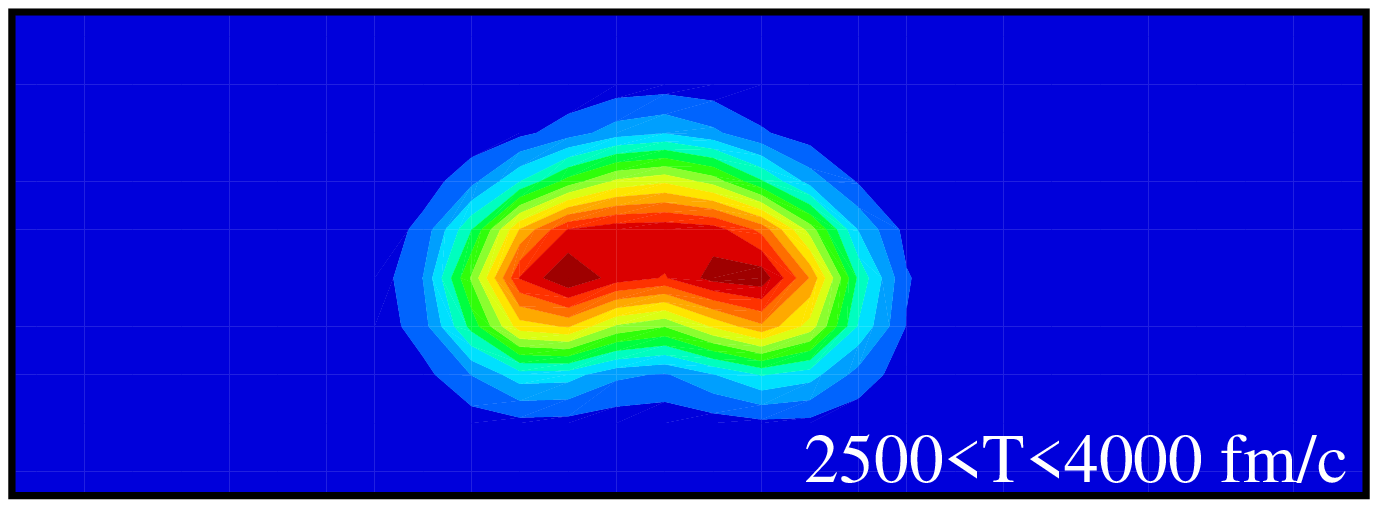}\\ \vspace{-0.03in}
\includegraphics*[scale=0.50]{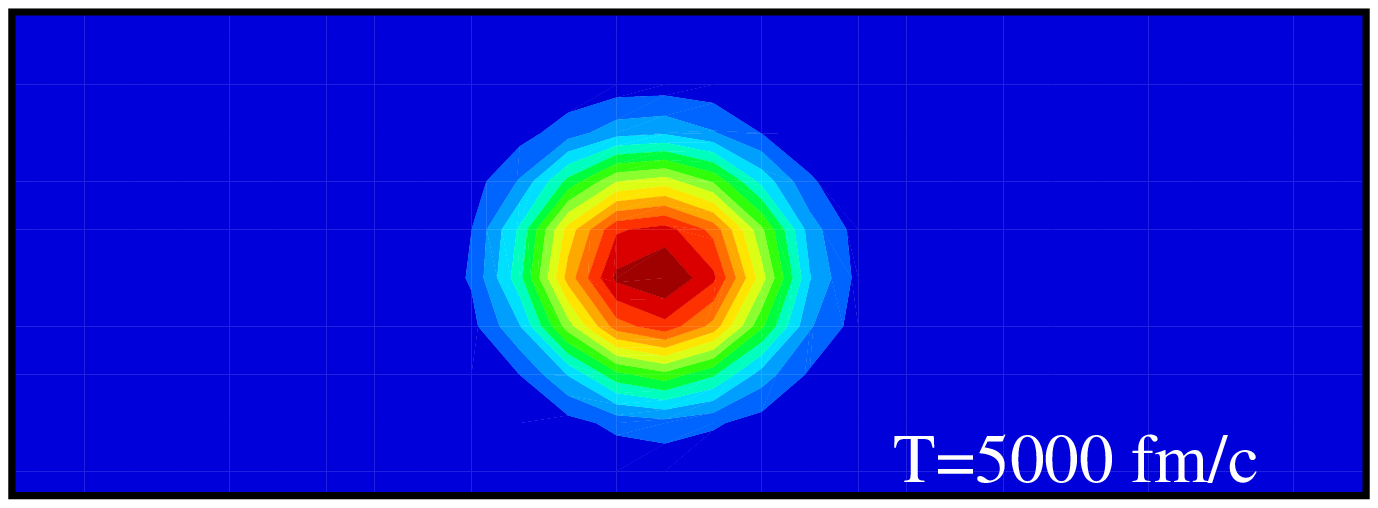}
\caption{\label{fig1}(Color online) Selected density profiles from TDHF time-evolution
of the $^{4}$He+$^{8}$Be head-on collision for initial
Be orientation angle $\beta=0^{\circ}$ using the SLy4 interaction.
The initial energy is $E_{\mathrm{c.m.}}=2$~MeV.}
\end{figure}

In our numerical calculations we have chosen a
Cartesian box which is $40$~fm along the collision axis and $24$~fm in
the other two directions.
Calculations are done in 3-D geometry and using the full Skyrme force (SLy4)~\cite{CB98}
as described in Ref.~\cite{UO06}.
We have tested these results for few cases by using a much larger numerical box and by
changing the time-step and found no appreciable difference.
Similarly, using different parametrizations of the modern Skyrme force all show the
same phenomenon.
We have chosen to study two different collisions leading to the linear chain configuration,
first the $^{4}$He+$^{8}$Be system and then the triple collision of three  $^{4}$He nuclei,
which may be astrophysically much less probable. The Hartree-Fock (HF) state for the
$^{8}$Be nucleus is axially symmetric. In Fig.~\ref{fig1} we show three snapshots from the
long time evolution of the $^{4}$He+$^{8}$Be collision at zero impact parameter and
$E_{\mathrm{c.m.}}=2$~MeV. The top panel of Fig.~\ref{fig1} shows the linear chain configuration
about which the system oscillates for times less than $2,500$~fm/$c$.
In particular, it
is remarkable that the moving clusters do not equilibrate while moving inside the
linear chain state
but retain their $2p-2n$ character, where one observes a complex
quasiperiodic motion with little damping up to this time.
Around $2500$~fm/$c$ the
system starts to bend and acquires a somewhat triangular shape as shown in the middle panel
of Fig.~\ref{fig1}.
The system still retains its cluster character with the center cluster moving off the reaction
plane cut shown in the figure, but can be clearly observed in volumetric three-dimensional
movies of the collision process.
The bending motion, where the center cluster oscillates somewhat perpendicular to the
left and right clusters continues for approximately $1000$~fm/$c$, with very little damping.
Finally, at even longer times the system relaxes into a relatively more compact
shape (bottom panel of Fig.~\ref{fig1}.).
Such mode changes, where the dynamical energy in the longitudinal direction is converted to a
transverse mode, while the system retains its cluster structure have never been seen in previous
TDHF calculations albeit this would not have been possible in calculations imposing axial symmetry.
Even in three-dimensional calculations, for an exactly central collision, the axial
symmetry would be preserved under ideal theoretical and numerical conditions. Therefore the
meaning of head-on collision ($b=0$) should be interpreted to have a small dispersion
around this value, which facilitates the mode change even for exactly central collisions.
\begin{figure}[!htb]
\includegraphics*[scale=0.38]{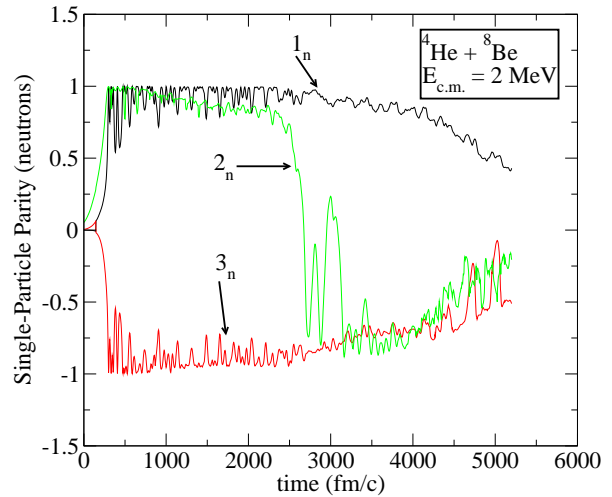}
\caption{\label{fig2}(Color online) Single-particle parities of the neutron states during the
collision of the $^{4}$He+$^{8}$Be system as a function of time at $E_{\mathrm{c.m.}}=2$~MeV.}
\end{figure}

We were also successful, for the first time, in creating a static
Hartree-Fock linear chain state orthogonal to the ground state in a
fully three-dimensional geometry.
This was achieved  by initializing one of the single-particle states to be
in the $s-d$ shell with positive parity rather than in the $p$ state with negative parity. This results in a linear chain
state similar to the one shown in the top panel of Fig.~\ref{fig1}. The fact that this state has two positive parity
and one negative parity state proves that it is exactly orthogonal to the $^{12}$C ground state which has one
positive parity and two negative parity states. 
A similar dependence on parity was also studied in cluster model calculations~\cite{IO06}.
In order to relate this observation to the dynamical mode changes
discussed above we have used the  density constraint method to calculate the potential energy and the
single-particle parities during the TDHF time-evolution.
In Fig.~\ref{fig2} we show the neutron single-particle parities as a function of collision time.
What is striking is that the combined system initially has the same parity signature as the
static linear chain state but at the time of bending one of the positive parity states starts
to decay towards negative parity and this decay continues as the system becomes closer to the
parity signatures of the ground state. Oscillations in the numerically calculated parities stem from
the fact that these are done during the dynamical evolution of the system.
The proton single-particle parities are almost exactly
the same as for the neutrons as anticipated.
\begin{figure}[!htb]
\includegraphics*[scale=0.38]{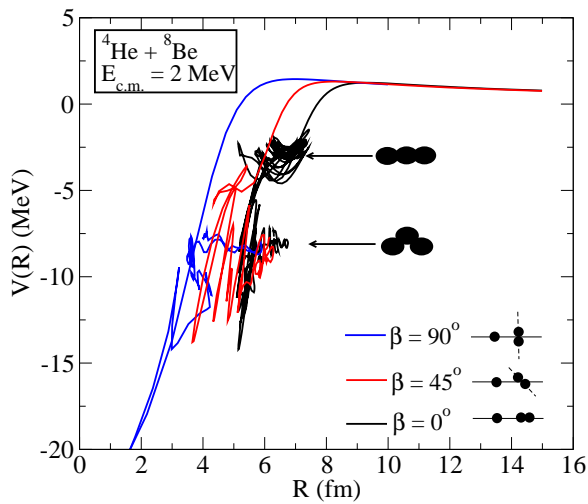}
\caption{\label{fig3}(Color online) Potential energy curves for the
collision of the $^{4}$He+$^{8}$Be system as a function of $R$ for three initial
alignments of the Be nucleus and at $E_{\mathrm{c.m.}}=2$~MeV.}
\end{figure}

In order to relate the observed mode changes more closely to the intrinsic energy of the system we
have also calculated the potential energy of the system as a function
of the ion-ion separation distance $R$~\cite{UO06b}. For the calculation of $R$  we have used the hybrid method
described in~\cite{UO09b}, which relates $R$ to the quadrupole moment of the system thus making it
possible to have a consistent definition of $R$ for large overlaps. 
The calculated potential energy curves are
shown in Fig.~\ref{fig3} as a function of three alignments of the $^{8}$Be nucleus with respect to the
collision axis labeled as angle $\beta$, and for the entire duration of the collision process.
Since for the real system the angular momentum of $^{8}$Be is projected all possible alignments
of the $^{8}$Be nucleus needs to be considered.
For all of the alignments the combined system climbs up a shallow potential barrier height of
approximately $1.24-1.44$~MeV, the lowest barrier being that of the $\beta=0^{o}$ potential,
thus making this alignment most probable under threshold conditions.
For the potential energy curve showing the head-on collision ($\beta=0^{o}$, black curve) we observe
that the system initially relaxes to a relatively shallow metastable minimum and oscillates about this
minimum until approximately $T=2500$~fm/$c$ at which points it slips down the curve towards the
second configuration as indicated by three alphas in a triangular configuration. After spending some
time in this configuration the system further slips down to even lower energy and more compact configuration.
The potential energy curves corresponding to $^{8}$Be initial alignment angles of  $\beta=45^{o}$ and $\beta=90^{o}$
(red and blue curves, respectively) undergo a different behavior, bypassing the linear chain minimum but
directly going to the triangular and subsequently to the compact configuration, the perpendicular
energy collision attaining the most compact and lowest energy configuration.
It is interesting to note that all of the potential energy curves spend some time in the triangular configuration.

\begin{figure}[!htb]
\includegraphics*[scale=0.38]{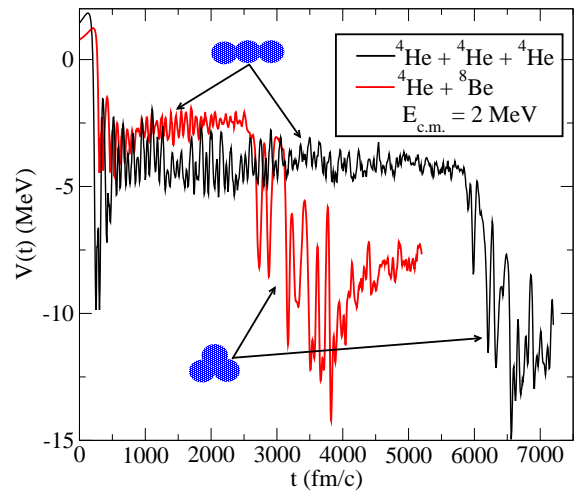}
\caption{\label{fig4}(Color online) Time development of the potential energy for the
head-on collision of the $^{4}$He+$^{8}$Be and the $^{4}$He+$^{4}$He+$^{4}$He systems
for $E_{\mathrm{c.m.}}=2$~MeV.}
\end{figure}

As an alternate collision leading to the same configuration we have also studied the collision of three $^{4}$He
nuclei, one at rest at the origin of the collision axis and the other two on each side boosted towards the center
with $1$~MeV energy. In Fig.~\ref{fig4} we contrast the time dependence of the potential energies for the
two different collisions. We observe that the three $^{4}$He collision process spends considerably longer time
(about $6000$~fm/$c$) undergoing quasiperiodic oscillations with very little damping
in the linear chain configuration before switching to bending and compact modes.

\begin{figure}[!htb]
\includegraphics*[scale=0.38]{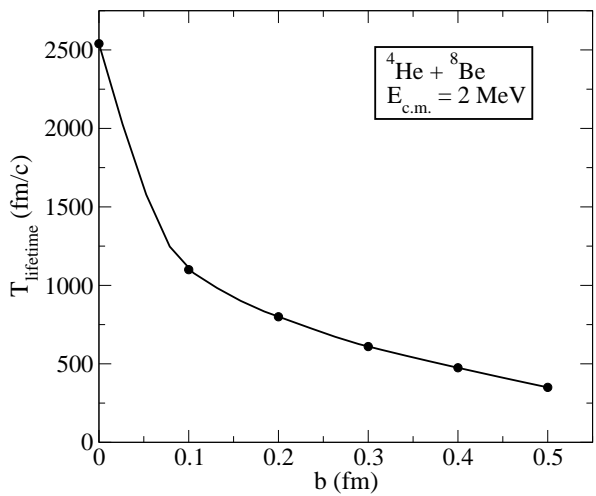}
\caption{\label{fig5} Time spent in the linear chain configuration as a function of the impact parameter $b$
for the $^{4}$He+$^{8}$Be system at $E_{\mathrm{c.m.}}=2$~MeV and $\beta=0^{o}$ alignment.}
\end{figure}

In order to gauge the stability of the linear chain configuration we have made systematic studies
as a function of impact parameter and center-of-mass energy, as well as a study using heavier Be
isotopes to determine the dependence on neutron number.
In Fig.~\ref{fig5} we show the dependence of the linear chain survival time on the impact parameter for the
$^{4}$He+$^{8}$Be system at $E_{\mathrm{c.m.}}=2$~MeV and $\beta=0^{o}$. We observe that as
the impact parameter increases the survival time rapidly decreases.
This decrease naturally
happens slower (faster) for lower (higher) energies. We have also done a similar study  for the time spent in the linear chain
configuration as a function of the center-of-mass energy for the $^{4}$He+$^{8}$Be system for $\beta=0^{o}$ alignment.
We decreased the energy in steps of $0.1$~MeV to find the lowest energy for which we form the linear chain configuration
(at lower energies the nuclei rebound due to Coulomb repulsion). At this energy of  $1.3$~MeV the
lifetime of the linear chain configuration increases to about $2875$~fm/$c$. As the energy is increased the lifetime
decreases gradually. 
In order to study the dependence of the linear chain state on the neutron number of the Be isotopes we
have repeated all of the above calculations using a $^{9}$Be nucleus instead. The calculations were
done by using all the time-odd terms in the Skyrme interaction appropriate for an odd-A nucleus.
While we do find an analogous
behavior in this study, the lifetime of the linear chain state is approximately $30$\% less than that of the
corresponding $^{8}$Be system.

The analysis of the long-time motion of collective nuclear phenomena in terms of their quasiperiodic
behavior is related to the long-time behavior of nonintegrable mechanical systems and addresses
questions as to the nature of energy dissipation and equilibration of energy.
In this work we have performed microscopic dynamical calculations of nuclear collisions to study
the formation of a metastable linear chain state of $^{12}$C.
The time evolution as calculated using the TDHF equations shows a characteristic quasiperiodic
exchange of alpha-like clusters in the density function corresponding to
a quasiperiodic motion along a static Hartree-Fock potential
energy surface, which is studied using the density constraint procedure.
We have shown that the calculations lead to the formation of a metastable linear chain state of three alpha-like clusters
which subsequently decays
to a lower energy triangular alpha-like configuration before acquiring a more compact shape.
This is the first observation of such mode changes in TDHF calculations and the results
seem to be analogous to the suggested astrophysical  mechanism for the
formation of $^{12}$C nuclei.

This work has been supported by the U.S. Department of Energy under grant No.
DE-FG02-96ER40963 with Vanderbilt University, and by the German BMBF
under contract No. 06F131. One of us (A.S.U.) thanks for the support by the Hessian
LOEWE initiative through the Helmholtz International Center for FAIR.

\end{document}